\documentclass[%
 aip,
 amsmath,amssymb,
 reprint,%
]{revtex4-1}

\usepackage{graphicx}
\usepackage{dcolumn}
\usepackage{bm}

\usepackage[utf8]{inputenc}
\usepackage[T1]{fontenc}
\usepackage{mathptmx}
\usepackage{etoolbox}
\usepackage{xcolor}
\makeatletter
\def\@email#1#2{%
 \endgroup
 \patchcmd{\titleblock@produce}
  {\frontmatter@RRAPformat}
  {\frontmatter@RRAPformat{\produce@RRAP{*#1\href{mailto:#2}{#2}}}\frontmatter@RRAPformat}
  {}{}
}%
\makeatother
\begin{document}
\preprint{AIP/123-QED}

\title[]{Direct experimental observation of total absorption and loss compensation using sound waves with complex frequencies}

\author{Anis Maddi}
\email{anis.maddi@univ-lemans.fr}
\affiliation{Laboratoire d’Acoustique de l’Université du Mans (LAUM), UMR 6613,  Institut d’Acoustique - Graduate School (IA-GS), CNRS, Le Mans Université, France.}

 \author{Gaelle Poignand}
\affiliation{Laboratoire d’Acoustique de l’Université du Mans (LAUM), UMR 6613,  Institut d’Acoustique - Graduate School (IA-GS), CNRS, Le Mans Université, France.}

 \author{Vassos Achilleos}
\affiliation{Laboratoire d’Acoustique de l’Université du Mans (LAUM), UMR 6613,  Institut d’Acoustique - Graduate School (IA-GS), CNRS, Le Mans Université, France.}

 \author{Vincent Pagneux}
\affiliation{Laboratoire d’Acoustique de l’Université du Mans (LAUM), UMR 6613,  Institut d’Acoustique - Graduate School (IA-GS), CNRS, Le Mans Université, France.}

\author{Guillaume Penelet}			
\affiliation{Laboratoire d’Acoustique de l’Université du Mans (LAUM), UMR 6613,  Institut d’Acoustique - Graduate School (IA-GS), CNRS, Le Mans Université, France.}


\date{\today} 

\begin{abstract}

In this study, we experimentally investigate the application of a transient signal with complex frequencies to the absorption and transmission of sound waves. Indeed, the emission of a wave with an exponentially varying amplitude in time is analogous, in the frequency domain, to a monochromatic wave with spatial gain or loss.  Our results show that by exciting a non-critically coupled Helmholtz resonator with a wave having a growing amplitude, total absorption can still be achieved. Furthermore, the lossy propagation of a traveling wave in a duct is also studied, and it is shown that the losses embedded in the complex wavenumber can be compensated by using a transient signal with decreasing amplitude. These results confirm the potential of complex frequency excitation for exploring new means of manipulating sound waves to mimic gain and loss.

\color{black}

\end{abstract}

\maketitle

\newpage
\section{Introduction}

\color{black}
The dissipation of acoustic energy is an inherent aspect of wave propagation which presents both challenges and opportunities for manipulating sound waves.  Once harnessed appropriately, this dissipation can lead to a controlled absorption of noise, which  constitutes a vast field of research. Recently, major advances have been achieved thanks to the progress in the field of acoustic metamaterials \cite{yang2017sound,gao2022acoustic,qu2022microwave,cummer2016controlling,zangeneh2019active,kumar2020recent}, notably for the design of sound wave absorbers using various types of resonant structures \cite{wang2024seven,huang2018acoustic,jimenez2016ultra,romero2020perfect,auregan2018ultra,huang2019acoustic,li2016acoustic,mei2012dark,yang2015subwavelength,jimenez2017quasiperfect}.

Sound wave absorption using acoustic metamaterials requires carefully engineered structures to obtain a high absorption. Among the prevalent systems employed are Helmholtz resonators, owing to their simplicity and scalability. In a reflection problem \cite{jimenez2016ultra,auregan2018ultra}, where a Helmholtz resonator is placed at the end of a waveguide, the full absorption of sound can be realized at the resonance frequency by  finely tuning the resonator \cite{romero2016perfect} to achieve critical coupling. While these types of structures are widely used and studied, they have predominantly been investigated in terms of real-frequency excitation and may not be suitable for addressing transient signals with amplitude growth or attenuation.

Recently, a growing interest in the use of transient signals with a complex frequency has emerged \cite{baranov2017coherent} to control waves. These excitations, characterized by temporal amplitude decay or growth, provide an effect that mimics the introduction of gain or loss, and have already exhibited promising features, including virtual absorption \cite{baranov2017coherent,krasnok2019anomalies,novitsky2023virtual,delage2022experimental,marini2022perfect}, virtual critical coupling \cite{ra2020virtual,hinney2024efficient}, $\mathcal{PT}$-symmetry \cite{li2020virtual,chen2024observation,trivedi2024anomalies} and loss compensation \cite{kim2023loss,guan2024compensating,guan2023overcoming}. This concept has already been implemented in various physical setups, including for electronic circuits \cite{li2020virtual,trivedi2024anomalies}, elastic waves \cite{trainiti2019coherent} and water waves \cite{euve2021transient}. In acoustics, virtual gain was recently used for the study of non-Hermitian topological systems \cite{gu2022transient} and the design of a superlens \cite{kim2023loss}. Aside from these two studies, experimental investigations in acoustics remain limited, with aspects like virtual sound absorption still to be explored.
\color{black}

\begin{figure*}[ht]
    \centering
    \includegraphics[width=179mm]{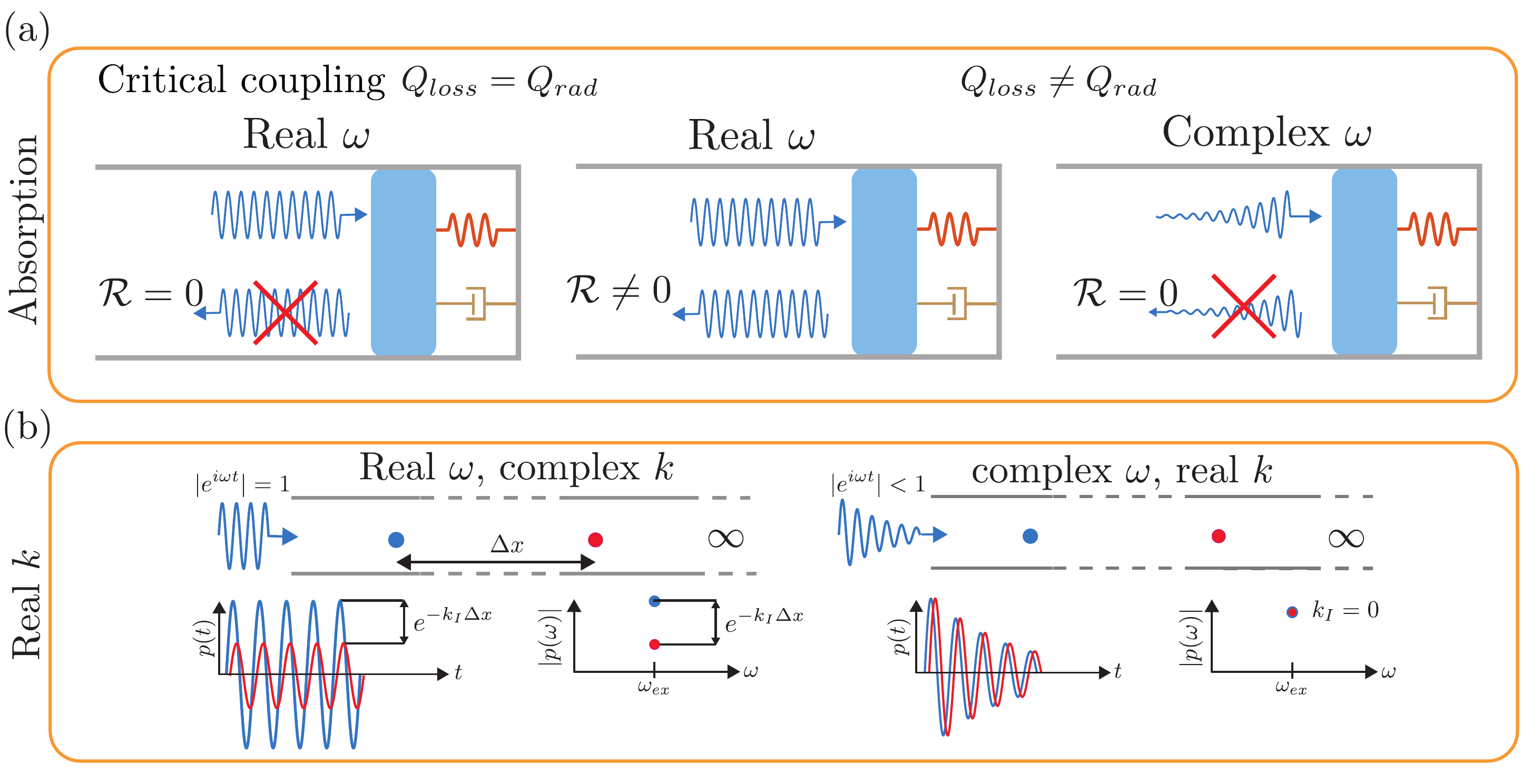}
    \caption{Illustrations of the problems investigated in this work \color{black} \color{black}.  (a) An absorption problem, where an acoustic resonator is placed at the end of a duct. The critical coupling condition ensures a perfect absorption for real frequencies (time convention $e^{i\omega t}$). When the resonator and duct are not critically coupled,  a total absorption of the incident waves can still occur when complex frequencies are used.
    (b) Propagation of sound waves in a dissipative media. The introduction of a complex frequencies allows to obtain a real wavenumber $k$ despite the losses.}
    \label{Fig_illus}
\end{figure*}

In this work, we experimentally investigate the use of complex frequencies for the absorption and transmission of airborne sound waves. Herein, two distinct problems are addressed experimentally. In the first part, the absorption of sound waves using a Helmholtz resonator excited with a complex frequency is investigated. The resonator is intentionally designed to be non-critically coupled when excited with real frequencies. However, when a transient signal in the form of an exponentially increasing (in time) signal is used as the input of the system, a full absorption of the incident waves is observed. In the second part, we demonstrate the possibility of obtaining a lossless transmission by counterbalancing the spatial decay embedded in the lossy wavenumber using the gain induced by an incident signal with a temporal decay \cite{guan2024compensating,kim2023loss}.

The plan of this paper is as follows. In section II, a brief theoretical description of an absorption problem using complex frequency is proposed where we show that a non-critically coupled resonator can still provide a full absorption of sound waves provided that the frequencies are extended to the complex domain (See Fig.\ref{Fig_illus}.(a) for a graphical illustration). An experimental investigation is then performed, with a simple system consisting of a Helmholtz resonator connected to an acoustic waveguide. A comparison between the measured temporal signals for real and complex frequency excitations is provided, and demonstrates an enhancement of sound wave absorption when the system is excited using waves with an exponentially growing amplitude. In section III, we show that the presence of complex frequencies also allow to compensate for the losses embedded in the wavenumber, in the extent that it allows  to obtain a real wavenumber despite the presence of viscothermal losses (See Fig.\ref{Fig_illus}.(b) for a graphical illustration).

\section{Transient sound absorption}

\subsection{Theoretical description}

\color{black}
 Helmholtz resonators are among the most commonly used resonators in acoustics, due to their simplicity and tunability. A Helmholtz resonator (HR) can be modeled as a lossy harmonic oscillator \cite{pierce1981introduction}, characterized by an acoustic impedance $Z_{r}$ given by (time convention $e^{i\omega t}$),

 \color{black}

\begin{equation}
    Z_r=\frac{K_r}{i\omega}\left[1-\frac{\omega^2}{\omega_0^2}  +i\frac{\omega}{\omega_0} \frac{1}{Q_{loss}}\right]
\end{equation}

where $\omega_0$, $Q_{loss}$ and $K_r$ represent respectively the resonance frequency, quality factor, and stiffness of the resonator.

\begin{figure}[h]
    \centering
    \includegraphics[width=0.9\linewidth]{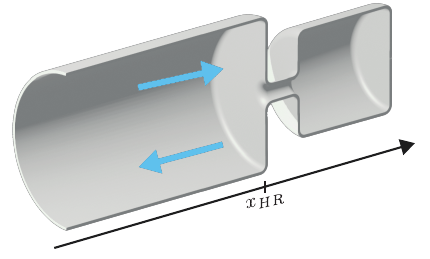}
    \caption{Reflection problem using a Helmholtz resonator connected to a duct at $x=x_{HR}$.}
    \label{fls}
\end{figure}

A commonly treated problem in acoustics is the one consisting of a resonator placed at the end of a waveguide to absorb sound waves (See Fig.\ref{fls} for an illustration using a HR). In this problem, the reflection coefficient is given by, 

\begin{equation}
    \mathcal{R}=\frac{1-\frac{\omega^2}{\omega_0^2}  +i\frac{\omega}{\omega_0}\left[ \frac{1}{Q_{loss}}-\frac{1}{Q_{rad}}\right]}{1-\frac{\omega^2}{\omega_0^2}  +i\frac{\omega}{\omega_0}\left[ \frac{1}{Q_{loss}}+\frac{1}{Q_{rad}}\right]},
\end{equation}
where $Q_{rad}=\frac{\omega_0}{Z_cK_r}$ represents the radiation factor, with $Z_c$ the characteristic impedance of the duct.

The optimal design of the absorbing material then amounts to reaching a critical coupling, which occurs by tuning the coupling between the waveguide and resonator so that $Q_{rad}= Q_{loss}$. Once this matching is achieved, perfect absorption with $\mathcal{R}=0$ is obtained at the HR resonance frequency $\omega=\omega_0$.

 Now, consider a similar absorption problem where the waveguide and resonator are not tuned to deliver a critical coupling, i.e, $Q_{rad}\ne Q_{loss}$. One question that may arise is whether or not it is still possible to fully absorb sound waves.

 To address this problem, one can expand the frequencies $\omega$ into the complex plane, such that  $\omega=\omega_R+i\omega_I$, in which the imaginary part describes a temporal growth ($\omega_I<0$) or decay ($\omega_I>0$) of the incident waves. When assuming a slow growth/decay  of the amplitude around the resonance frequency (i.e $\omega_0 \gg\vert \omega_I\vert$), the reflection coefficient  simplifies to,

  \begin{equation}
      \mathcal{R}\approx\frac{1-\frac{\omega_R^2  }{\omega_0^2}  -\frac{\omega_I}{\omega_0}\left[\frac{1}{Q_{loss}}-\frac{1}{Q_{rad}}\right]+i\frac{\omega_R}{\omega_0}\left[ \frac{1}{Q_{loss}}-\frac{1}{Q_{rad}}-\frac{2\omega_I}{\omega_0} \right]}{1-\frac{\omega_R^2  }{\omega_0^2}  -\frac{\omega_I}{\omega_0}\left[\frac{1}{Q_{loss}}+\frac{1}{Q_{rad}}\right] +i\frac{\omega_R}{\omega_0}\left[ \frac{1}{Q_{loss}}+\frac{1}{Q_{rad}}-\frac{2\omega_I}{\omega_0} \right]}.
  \end{equation}
\color{black}

The zero of reflection can be found by solving simultaneously the real and imaginary parts of the numerator of Eq.(3) leading to the following approximation of the real and imaginary part of the angular frequency
 \begin{align}
     \omega_I &\approx\frac{\omega_0}{2}\left[\frac{Q_{rad}-Q_{loss}}{Q_{loss}Q_{rad}}\right].\\
     \omega_R &\approx \omega_0
     \label{Eq.5}
 \end{align}

Consequently, even if $Q_{rad}\ne Q_{loss}$ a perfect absorption around the  resonance frequency is possible by adjusting the imaginary part of the frequency $\omega_I$.  Starting from Eq.(\ref{Eq.5}), three coupling regimes can be defined:

\begin{itemize}
    \item Under-coupled system ($Q_{loss}>Q_{rad}$): A perfect absorption of the incident waves can be realized using a signal with exponentially increasing amplitude $\omega_I<0$. 
    \item Critically coupled system ($Q_{loss}=Q_{rad}$): This case corresponds to the traditional harmonic regime, where it was demonstrated \cite{romero2016perfect,romero2020design,jimenez2016ultra,huang2023sound} that a perfect absorption can be realized using real frequencies  $\omega_I=0$ .
    \item Over-coupled system ($Q_{loss}<Q_{rad}$): A perfect absorption of the incident waves can be realized using a signal with exponentially decaying amplitude $\omega_I>0$. 
\end{itemize}

While all three regimes above can be investigated experimentally, we decided in the scope of this work to only explore the under-coupled configuration.

Note that for an under-coupled system, the intrinsic losses of the HR are smaller than the radiation loss. It appears that the presence of the complex frequency serves to introduce further virtual losses to match the two factors \cite{li2020virtual}.  Once the exponential growth is adjusted accordingly to balance the radiative and intrinsic losses, a virtual critical coupling is obtained.

\subsection{Experimental Setup}

\begin{figure*}[ht]
    \centering
\includegraphics[width=172mm]{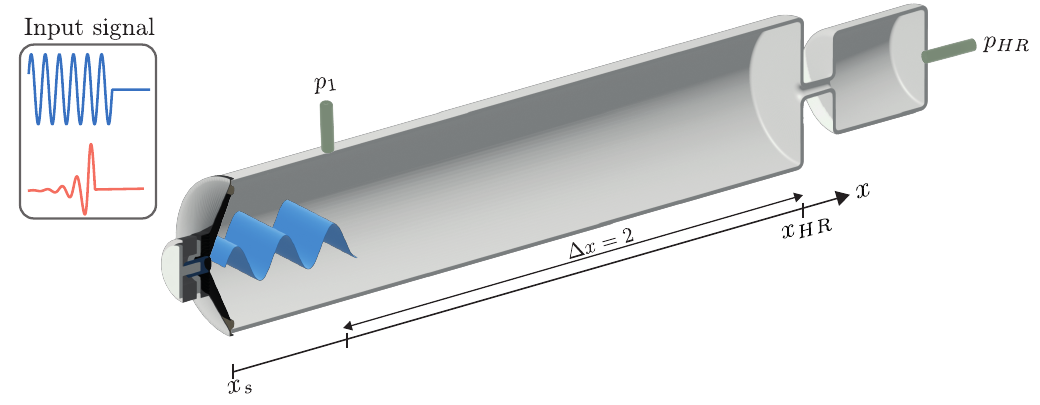}
    \caption{Sketch of the experimental apparatus. A non-criticallty coupled Helmholtz resonator is placed at right end of a duct (at $x=x_{HR}$). The system is excited using a source placed at the left end ($x=x_s$), which drives the system with a real or complex frequency. Two microphones are used in this experiment. The first measures the pressure $p_1$ at a distance $\Delta x=2$m from the HR entrance, while the second measures the pressure inside the resonator $p_{HR}$.
    }
    \label{Setup}
\end{figure*}

In this section,  the absorption of an under-coupled Helmholtz resonator excited with an exponentially growing time signal is investigated experimentally.

\begin{figure}[h]
    \centering
\includegraphics[width=0.48\textwidth]{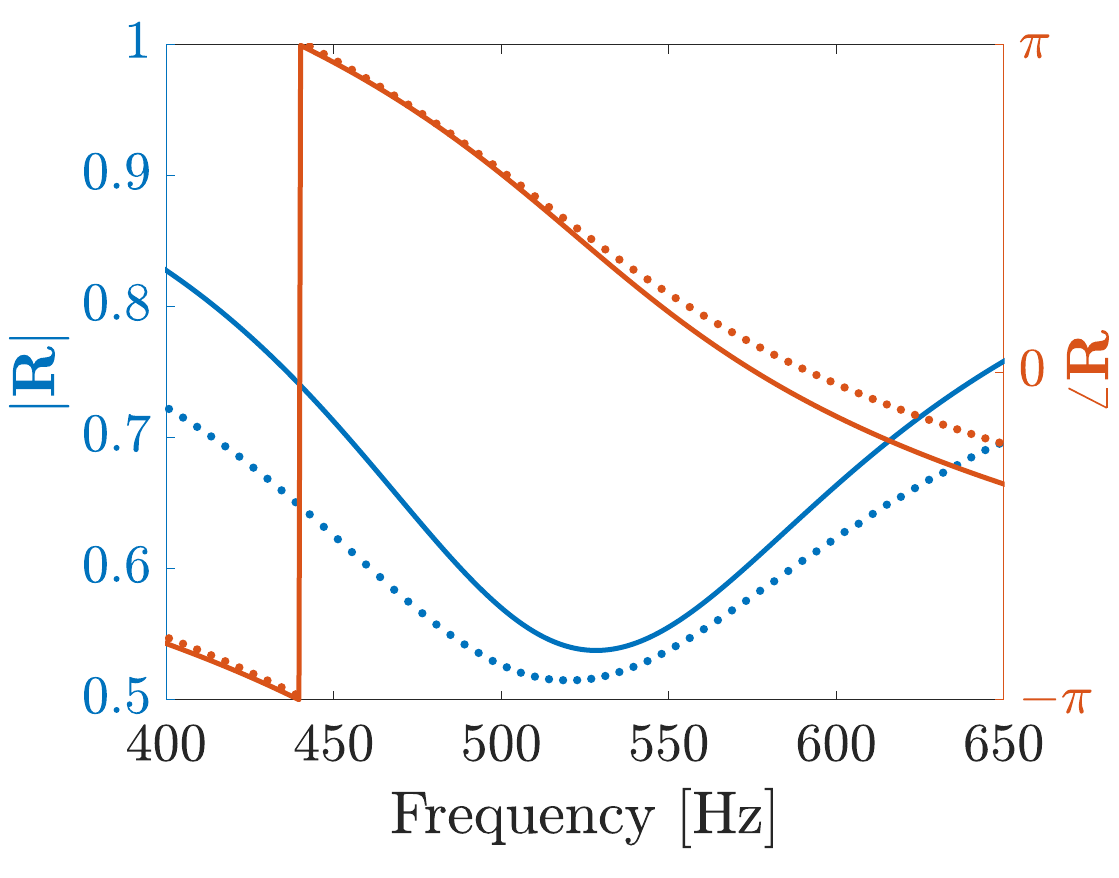}
    \caption{Magnitude and phase of the reflection coefficient of the Helmholtz resonator as functions of the frequency. Dots and lines represents the experimental and theoretical results, respectively. }
    \label{HR Ref}
\end{figure}

Figure \ref{Setup} depicts the experimental setup used in this study to investigate sound absorption using transient signals with amplitude growth. The system consists of a circular duct of  internal radius $R=4$ mm  and length $L=2.15$ m connected to a Helmholtz resonator at the right end ($x=x_{HR}$), and to a loudspeaker (Visaton, K50WP) at the left end ($x=x_s$). The Helmholtz resonator is designed so that the coupling with the waveguide corresponds to the under-coupled regime. When excited in the traditional harmonic regime (real frequencies), the HR does not fully absorb the impinging sound waves. However, when a signal with an adequate exponential amplitude growth is employed, then total absorption should be achieved.  A microphone (Bruel \& Kjaer, model 4938) is placed inside the resonator to measure the pressure $p_{HR}$, while a second microphone $p_1$ is placed close to the acoustic source, at a distance $\Delta x=2$ meters from the resonator entrance $x_{HR}$.

\begin{figure*}
    \centering
    \includegraphics[width=138mm]{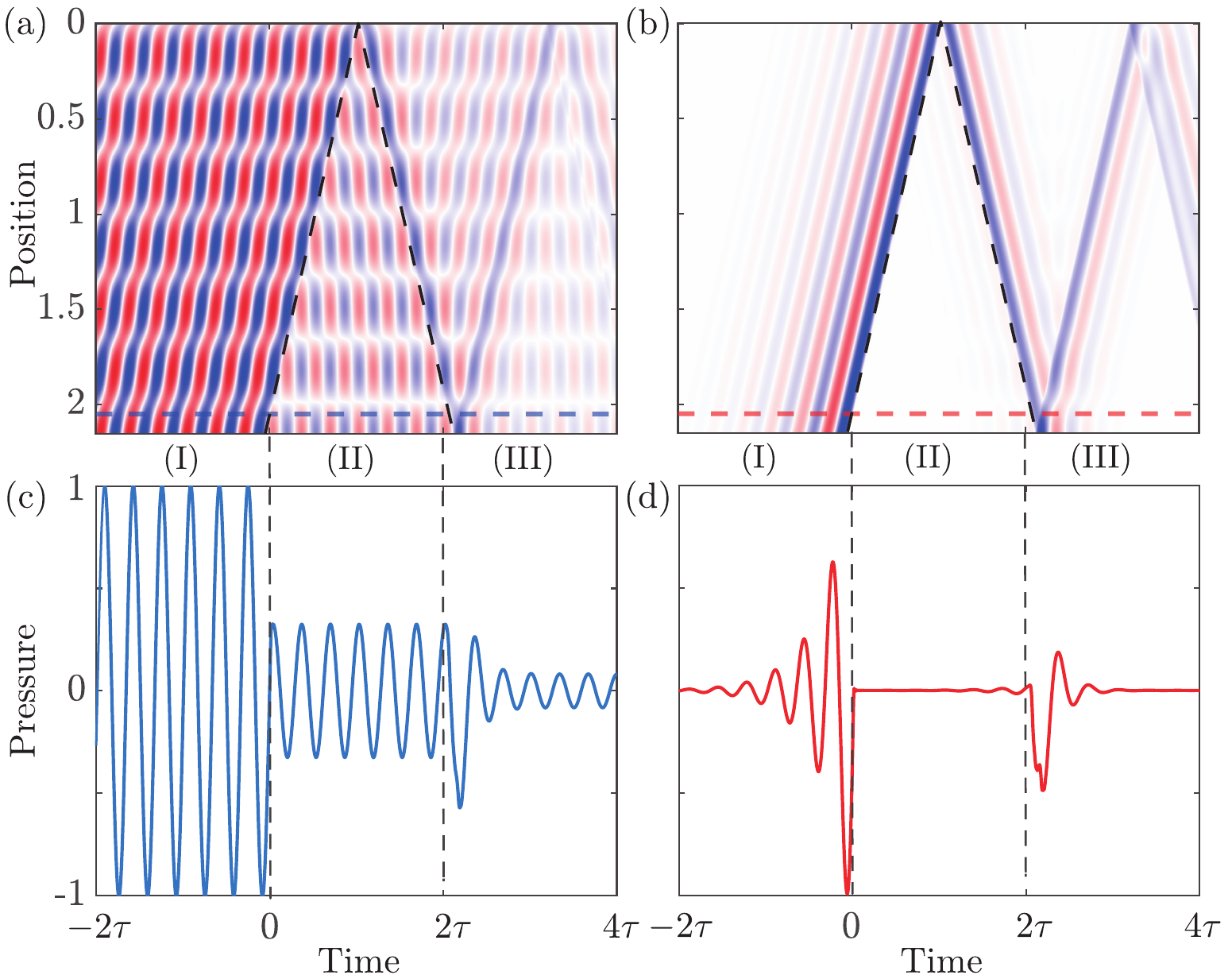}
    \caption{Numerical results using a 1D FDTD to illustrate the principle of the experimental apparatus. The damped oscillator is placed at $x_{HR}=0$ and the source at $x_s=2.15$.  (a-b) Spatiotemporal evolution of the pressure for a system excited using a real frequency $\omega_I=0$ and a complex frequency $\omega_I=\Im(\omega_z)$. (c-d) correspond to the pressure evolution as functions of the time ($\tau=\Delta x/c$) for a fixed position highlighted by the blue and red discontinuous lines in (a-b) (i.e. at a distance $\Delta x=2$m from the oscillator).}
    \label{FDTD}
\end{figure*}

  The magnitude and phase of the reflection coefficient of the HR at position $x_{HR}$ are plotted in Fig.\ref{HR Ref} as functions of real frequencies. The dotted and solid lines represent, respectively, the experimental and numerical results obtained using a transfer matrix approach. As discussed previously, the HR reflects a portion of the impinging waves,  as highlighted by the magnitude of the reflection coefficient which reaches $0.53$ around its resonance frequency.  The pole and zero of the reflection coefficient $\mathcal{R}$ can be extracted from the experimental results by using a rational polynomial fitting algorithm \cite{richardson1982parameter} (please see Supplementary information 1 for more details on the fitting).

The zero of the reflection coefficient corresponds to the complex frequency $\omega_{z}$ that allows to fully absorb the incident waves, i.e. $\mathcal{R}(\omega_z)=0$. 
In the present case, the fitting of the experimental results indicates a zero of reflection at the complex frequency $\omega_z/(2\pi) \approx 514-75i$ Hz (See Supplementary information 1). This value is also close to the one obtained numerically using a transfer matrix method with complex frequencies input.

In the following, the acoustic source is placed at $x=x_s$ (see Fig.\ref{Setup}) and serves to excite the system with the following input, 

\begin{align}
        p_{in}&=A\sin\left[\Re(\omega_z)t\right]e^{-\omega_I t}, \, t\le t_c \\
    p_{in}&=0 , \, t>t_c.
\end{align}

where $A$ and $t_c$ are respectively the initial amplitude of the signal and the cut-off time. The cut-off time $t_c$ represents the time for which the transient excitation is stopped. In the scope of this study, the imaginary part is set to two values: $\omega_I=0$ and $\omega_I=\Im(\omega_z) $, corresponding respectively to the traditional harmonic regime (constant amplitude) and an exponentially growing amplitude where the growth rate is adjusted to achieve total absorption.

\paragraph*{Numerical results}

Before addressing the experimental results, in order to understand the design of the setup, we first briefly discuss some numerical results obtained using a one-dimensional Finite-Difference Time-Domain (1D FDTD) method for a Helmholtz resonator coupled to an acoustic duct. In this model which is used for an illustration purpose, the HR is represented as a damped oscillator.

Figure \ref{FDTD}.(a-b) provides a spatiotemporal representation of the pressure, where the horizontal and vertical axes correspond to the temporal and spatial coordinates, respectively. This visualization allows to split the problem into three distinct regions:

\begin{enumerate}
    \item Region (I) represents the pressure inside the waveguide before stopping the excitation, which corresponds to the sum of left- and right-going waves for a real-frequency excitation (a), and to a right-going wave for a complex frequency excitation (b)
    \item Region (II) takes place after halting the excitation of the system, such that the signal comprises predominantly the waves reflected from the HR. 
    \item Region (III) is  a combination of the free response of the HR and of the waves (remaining in II) traveling back and forth between the left side (i.e., the source which acts act as a rigid termination when switched off) and the right side (HR).
\end{enumerate}

The primary region of interest in this study is Region (II), which forms a triangular shape that expands as the distance from the resonator (placed at $x_{HR}=0$) increases.

Figure \ref{FDTD}.(a) represents the system where the non-critically coupled oscillator is excited by a real frequency, and a non-vanishing amplitude of pressure is observed in region (II) due to the reflection by the HR. In contrast, Fig.\ref{FDTD}.(b) illustrates the same oscillator excited using a complex frequency with an exponentially growing amplitude ($\omega_I=\Im(\omega_z)$). In this case, the amplitude vanishes, demonstrating the absorption of sound waves. Note that the further the pressure is observed from the HR inlet, the longer it takes to observe the reflected waves (Reg. II) , making virtual absorption easier  to visualize. \color{black}

In the experiments presented in the next section, a microphone is placed to measure the pressure at a distance of  $\Delta x=2$m from the oscillator. This placement ensures that a sufficient number of oscillations can be observed before releasing the stored energy (region III). Figure \ref{FDTD}.(c-d) displays the numerical results using the 1D FDTD model  at $\Delta x=2$ m from the HR entrance as functions of time. As discussed previously, the total absorption of sound waves can be observed in region (II) when the system is excited using a complex frequency.

\subsection{Experimental results}

In the following, the experimental results obtained using the apparatus highlighted in Fig.\ref{Setup} are discussed, where it is shown that a total absorption of incident waves can still be realized using a non-critically coupled resonator by realizing a virtual critical coupling.

\begin{figure*}[ht]
    \centering
\includegraphics[width=172mm]{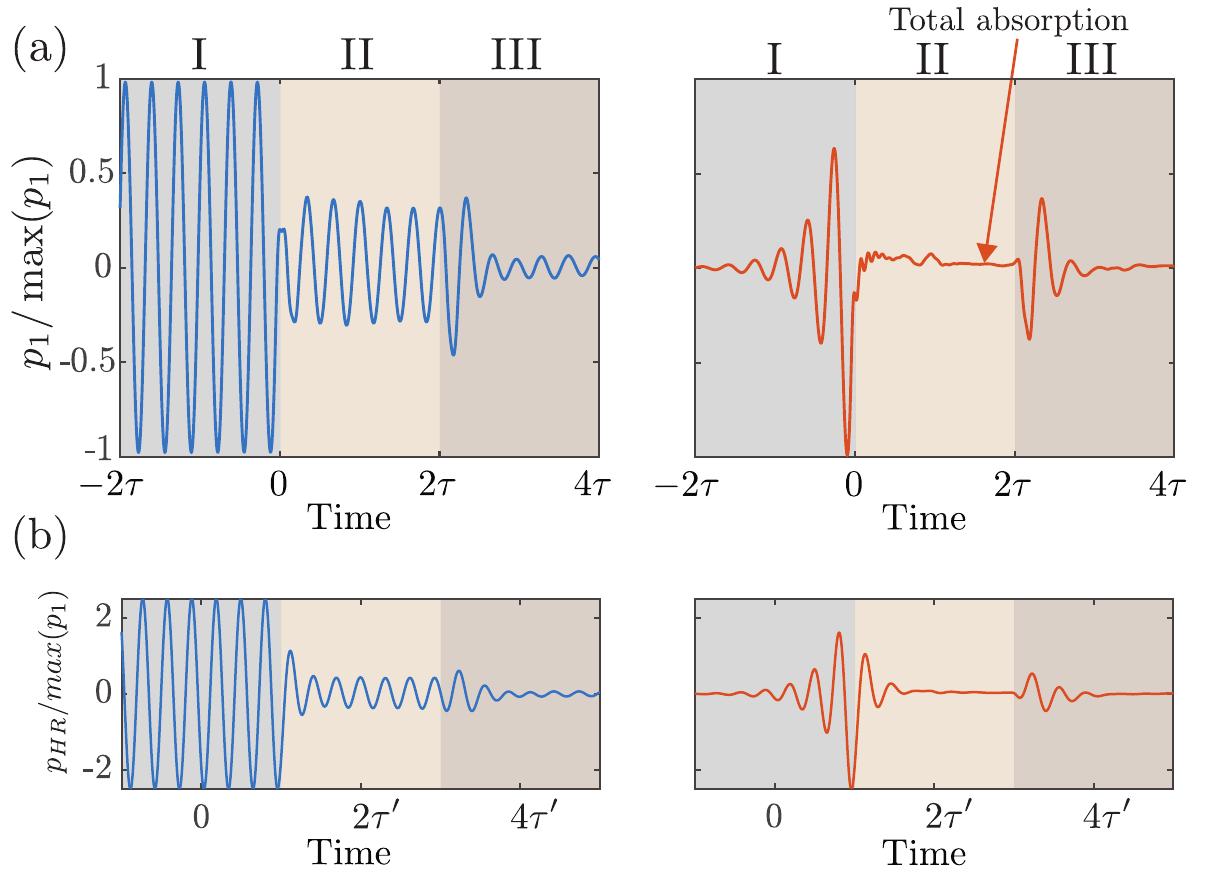}
    \caption{ Experimental results. (a) The normalized pressure  measured by the microphone $p_1$ as functions of time, where $\tau=\Delta x/c$ . (b) The normalized pressure  measured inside the Helmholtz resonator  (microphone $p_{HR}$) as functions of time, where $\tau'=L/c$. }
    \label{Res.Exp}
\end{figure*}

Figure \ref{Res.Exp}.(a) depicts the normalized pressures $p_1$ measured close to the acoustic source as functions of time, where the blue and red colors represent the results obtained either with a real frequency excitation or with a complex frequency excitation, respectively. As discussed previously, each signal can be divided into three distinct segments:

\begin{enumerate}
    \item Region (I) represents the measured signals before the cutoff time  at $t_c=0$.
  \item Region (II) happens after the excitation of the system is halted, during which the signal primarily consists of waves reflected from the Helmholtz resonator (HR). The total duration of this phase is \( 2\tau \), corresponding to twice the time of flight between the HR and the source. This region enables to highlight the absorption of the system.
    \item (III) represents the free-response of the oscillator due to the halted excitation, in addition to any reflected waves from the source and HR.
\end{enumerate}

The blue line represents an impinging signal with a constant amplitude, i.e.  $\omega_I$=0, where the normalized amplitude is unitary before the cut-off time (Region I).
Since no more incident wave is provided in Region II,  only the reflected waves remain traveling back and forth in the system. The signal has a lower amplitude estimated at $0.36$ as compared to the signal before the cut-off. 
 Once the excitation is halted, the HR starts free oscillations around time $t=2\tau$, and radiates the stored energy with a strong temporal decay. 

Next, the pressure measured for an exponentially growing signal in time that yields zero reflection ($\omega_I/(2\pi)=-75$ Hz) is displayed in red. In contrast to the previous case, a signal with a small amplitude after the cutoff time (Region II) is observed, highlighting the presence of a quasi-total absorption.  In Region (III), the energy stored in the resonator is released with an exponentially decaying amplitude, as discussed previously.  This is followed by  back-scattered waves, which result from reflection  caused by the source at the opposite end of the system.

A different perspective is provided in Fig.\ref{Res.Exp}.(b), which depicts the normalized pressure inside the resonator as functions of time. Before the cut-off (as observed by the microphone around $t=\tau'$), the pressure inside the resonator is higher than that near the acoustic source. When the acoustic excitation is stopped, the system starts releasing the accumulated pressure (around time $t=\tau'$). Then, between $t=\tau'$ and $t=3\tau'$, the remaining waves in the waveguide are sensed and excite the HR again. Once this excitation stops around $t=3\tau'$, the accumulated pressure is released again. \color{black} It is observed from the temporal measurement between $\tau'$ and $3\tau'$ that the signal with a complex frequency (red line) has a vanishing amplitude, which indicates an absorption of the sound waves. Note that this perspective corresponds to the horizontal line at the position $x=0$ in Fig.\ref{FDTD}.(a). \color{black}

Interestingly, the release of the stored energy is similar whether the system is excited with real or complex frequencies [see the beginning of region.(III) in Fig.\ref{Res.Exp}.(a) or (II) in Fig.\ref{Res.Exp}.(b)]. Note that for each excitation, the feeding time and amplitude before the cut-off are similar. Additionally, this release seems to be associated with the pole of the reflection coefficient $\mathcal{R}$ \cite{delage2022experimental}.

.

\section{Loss compensation}

In the previous section, we demonstrated that exponentially growing impinging waves allow for the introduction of additional losses that can be used to tune the absorption of a Helmholtz resonator. It is indeed possible to use a signal with decaying amplitude to introduce virtual gain in an acoustic system. In the following section,  the use of complex frequencies is explored as a way to introduce compensating gain during the propagation of sound waves, such that the spatial decay due to the lossy wavenumber is balanced using the complex frequency \cite{guan2024compensating,kim2023loss}.

The propagation of forward right-going traveling plane waves  between two spatial coordinates $x_1$ and $x_2$ satisfies the relation,

\begin{equation}
    p_{2}(\omega)= e^{-ik(x_2-x_1)}p_{1}(\omega).
          \label{Eq.10}
\end{equation}

where $k$ denotes the lossy wavenumber, which takes into account the thermo-viscous losses. In the case of a wide duct \cite{pierce1981introduction}, the wavenumber can be written in the following form,

\begin{equation}
      k=\frac{\omega}{c}+(1-i)\beta,
      \label{Eq.11}
\end{equation}

where $\beta(\omega)$ is a real-valued term that accounts for the losses. For simplicity, this term will be written in the form of $\beta(\omega)=a\sqrt{\omega}$, where $a>0$ is a positive constant.

Equations \ref{Eq.10} and \ref{Eq.11} imply that for any real angular frequency $\omega$, the propagation of waves along a duct of length $\Delta x=x_2-x_1>0$ results in an exponential amplitude decay given by $e^{-\beta \Delta x}$.  In the following, a complex frequency in the form of $\omega=\omega_R+i\omega_I$  is introduced in the description of the wavenumber $k$ (Equation \ref{Eq.11}). Assuming that the oscillating component is larger than the decay rate, i.e. $\omega_R>>\omega_I$, Equation \ref{Eq.11} simplifies to,

\begin{multline}
k = \frac{\omega_R}{c_0} + a\sqrt{\omega_R} \left( 1 - \frac{\omega_I}{2\omega_R} \right) \\
+ i\left[ \frac{\omega_I}{c_0} + a\sqrt{\omega_R} \left( \frac{\omega_I}{2\omega_R} - 1 \right) \right]
\end{multline}

Finally, by finding the solutions that cancel the imaginary parts of the wavenumber, i.e. $\Im(k)=0$, the following condition on the imaginary part of the complex frequency is obtained,

\begin{equation}
\omega_I = \underbrace{\frac{2a c_0 {\omega_R}}{2\sqrt{\omega_R} + a c_0 }}_\epsilon
\end{equation}
Hence, when the imaginary part of the frequency is adjusted, such that $\omega_I=\epsilon$, the wave number $k$ becomes real. As a result, a virtual lossless wave propagation is achieved, with $ \vert p_{1} \vert= \vert p_{2}\vert$. Moreover, since $\epsilon>0$, the resulting incident signal  always has a decreasing amplitude.

\begin{figure}
    \centering
\includegraphics[width=78mm]{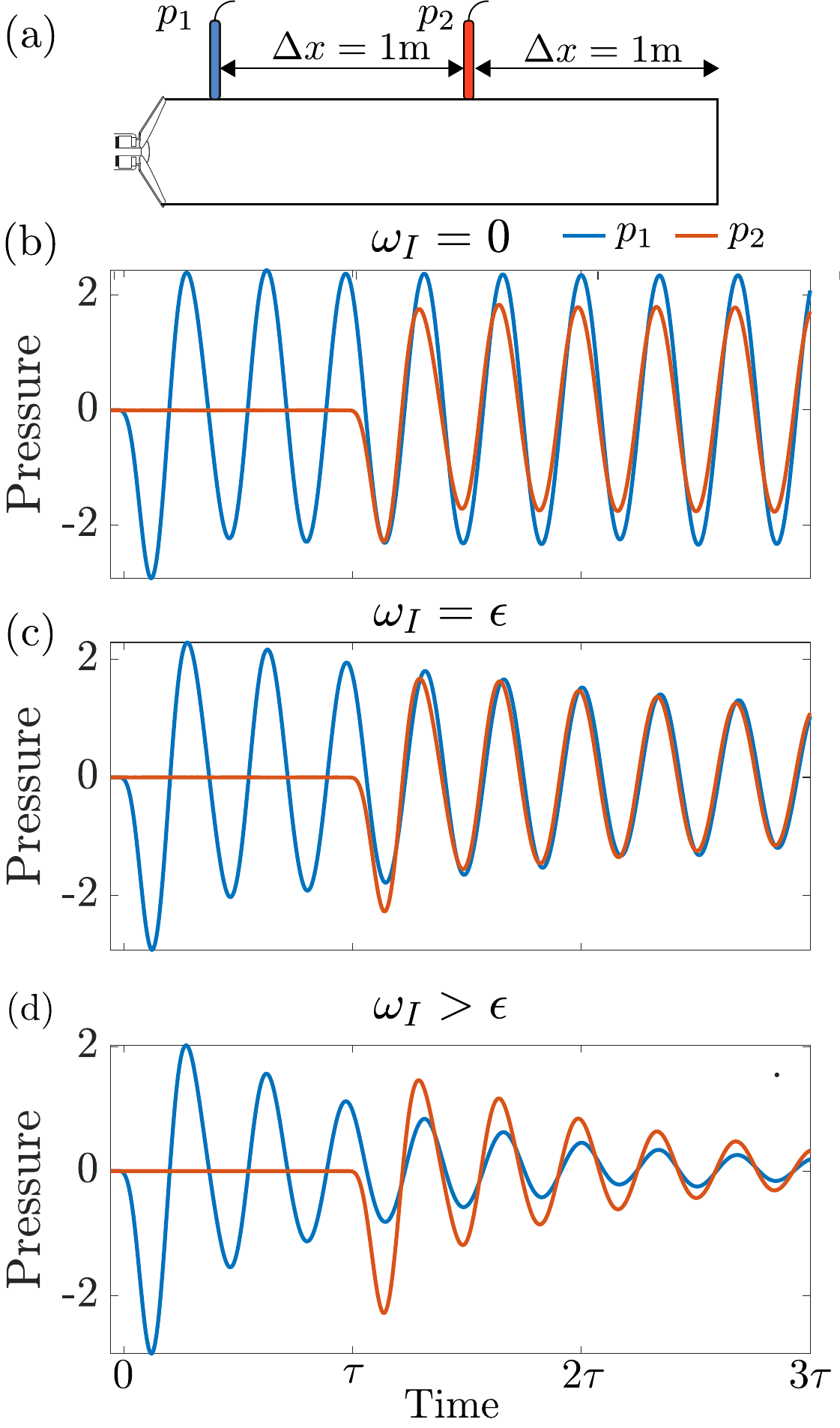}
    \caption{Experimental apparatus and results. (a) Experimental apparatus that consists of a duct, in which two microphone are flush-mounted and spaced by a distance $\Delta x =1$m. The system is excited with signals that have an oscillating frequency of $\omega_R/(2\pi)=1000$ Hz. (b,c,d)  Pressure measured as functions of time for different $\omega_I$. (b) Traditional harmonic regime $\omega_I=0$, (c) Loss compensation $\omega_I=\epsilon$, (d) over-compensation $\omega_I>\epsilon$. }
    \label{losless}
\end{figure}

Figure \ref{losless}.(a) illustrates the experimental apparatus used to demonstrate the spatial loss compensation using the gain embedded in the complex frequency. The system consists of a duct (of $4$ mm radius and $2.15$ m length) with a rigid termination at the right end, and a source placed at the left end. Two microphones are flush mounted along the duct, the first close to the source, and the second one placed at an equidistance of $ \Delta x =1$ m from the first microphone and the rigid end. In the following, the system is excited by complex frequencies, with a constant real part $\omega_R/(2\pi)=1000$ Hz, and a varying imaginary part $\omega_I$. Herein, only the time slot where the contribution of the rigid plug can be ignored is considered.   This corresponds to the propagation time of $3\tau$, such that $\tau=\Delta x/c$ is required for the signal to reach the microphone at the center, and $2\tau$ for a forward and backward propagation toward that same microphone.

Figure \ref{losless} depicts the experimental acoustic pressure measured by the two microphones as functions of time, for three distinct imaginary parts. The signals are shown here only in the time interval preceding the reflection of the waves by the rigid wall at the end of the duct. When the incident waves consist of a purely real frequency (panel a), i.e. $\omega_I=0$, the microphone close to the source (blue) measures a higher amplitude than the microphone placed $\Delta x$ away (red), due to thermoviscous losses $\Im(k)<0$. However, when the decay rate is tuned such that the wave number becomes purely real (panel b), the pressures measured have equal amplitudes. Finally, when the imaginary part is set above the threshold (panel c), the signal measured at the farthest microphone has a higher amplitude than that near the source, highlighting an overcompensation of the losses characterized by $\Im(k)>0$.


Note that the choice of the frequency $\omega_R$ was arbitrary. For any frequency $\omega_R$, there exists a corresponding $\omega_I$ that compensates for viscothermal losses via virtual gain. Incidentally, due to the real part of the frequency $\omega_R$ and the distance between the two microphones $\Delta x$, the two signals are almost in phase.

\color{black}

\section{Conclusion}

In this work, a simple experimental test-bench is used to demonstrate that the concept of virtual gain can be easily applied to airborne sound waves. More precisely, the use of transient excitations with exponentially decreasing or increasing amplitude is investigated to achieve perfect absorption by a (non-critically coupled) Helmholtz resonator, or for the compensation of spatial losses induced by a complex wavenumber.

First, we show that by using a non-critically coupled Helmholtz resonator, incident waves with a growing amplitude mimic the introduction of additional losses, leading to total absorption despite the impedance mismatch. This absorption is \color{black} directly \color{black} observed experimentally by analyzing the pressure measured (after halting the excitation) by a microphone placed far from the resonator. Secondly, we demonstrate that incident waves with decaying amplitude introduce gain to the system, enabling the compensation of inherent viscothermal losses embedded in the complex wavenumber. This is investigated experimentally by measuring the pressure at two different spatial positions, where similar amplitudes are observed when the imaginary part is appropriately adjusted.

The experimental findings presented in this work confirm the potential of complex frequency excitation is applicable for the control of airborne sound waves. Such excitation indeed provides an interesting way to introduce gain and loss in passive systems, and it could be employed in more complex scenarios.

\color{black}

\section{Acknowledgments}
We would like to thank Cyril Desjouy for his valuable contribution to the FDTD simulation.

V.A. acknowledges financial support from the NoHeNA project funded under the program Etoiles Montantes of the Region Pays de la Loire. V.A. is supported by the EU H2020 ERC StG ”NASA” Grant Agreement No. 101077954

\bibliography{sampbib}

\end{document}